%% file: 00_main.tex
\documentclass[screen, acmsmall, sigconf]{acmart}
\AtBeginDocument{%
  }

\usepackage{enumitem}
\usepackage{microtype}

\definecolor{editCol}{rgb}{0.0, 0.0, 0.0}
\newcommand{\edit}[1]{{\textcolor{editCol}{#1}}}

\usepackage{verbatim}

\copyrightyear{2026}
\acmYear{2026}
\setcopyright{cc}
\setcctype{by}
\acmConference[IDC '26]{Proceedings of the 25th Interaction Design and Children Conference}{June 22--25, 2026}{Brighton, United Kingdom}
\acmBooktitle{Proceedings of the 25th Interaction Design and Children Conference (IDC '26), June 22--25, 2026, Brighton, United Kingdom}
\acmDOI{10.1145/3773077.3806116}
\acmISBN{979-8-4007-2283-7/2026/06}
\begin{document}

\title[Participation and Power]{Participation and Power: A Case Study of Using Ecological Momentary Assessment to Engage Adolescents in Academic Research}


\author{Ozioma C. Oguine}
\orcid{0000-0003-2434-1400}
\affiliation{%
\department{Computer Science and Engineering}
  \institution{University of Notre Dame}
  \city{Notre Dame}
  \state{Indiana}
  \country{USA}}
\email{ooguine@nd.edu}

\author{Elmira Rashidi}
\orcid{0009-0003-2206-7433}
\affiliation{%
\department{Luddy School of Informatics, Computing, and Engineering,}
  \institution{Indiana University}
  \city{Indianapolis}
  \state{Indiana}
  \country{USA}}
\email{erashidi@iu.edu}

\author{Pamela J. Wisniewski}
\orcid{0000-0002-6223-1029}
\affiliation{%
\department{Socio-Technical Interaction Research Lab}
  \institution{International Computer Science Institute}
  \city{Berkeley}
  \state{California}
  \country{USA}}
\email{pwisniewski@icsi.berkeley.edu}

\author{Karla Badillo-Urquiola}
\orcid{0000-0002-1165-3619}
\affiliation{%
\department{Computer Science and Engineering}
  \institution{University of Notre Dame}
  \city{Notre Dame}
  \state{Indiana}
  \country{USA}}
\email{kbadillou@nd.edu}

\renewcommand{\shortauthors}{Ozioma C. Oguine et al.}

\begin{abstract}
Ecological Momentary Assessment (EMA) is widely used to study adolescents’ experiences; yet, how the \textit{design} of EMA platforms shapes engagement,  research practices, and power dynamics in youth studies remains under-examined. We developed a youth-centered EMA platform prioritizing youth engagement and researcher support, and evaluated it through a case study on a longitudinal investigation with adolescent twins focused on mental health and sleep behavior. Interviews with the research team examined how the platform design choices shaped participant onboarding, sustained engagement, risk monitoring, and data interpretation. The app’s teen-centered design and gamified features sustained teen engagement, while the web portal streamlined administrative oversight through a centralized dashboard. However, technical instability and rigid data structures created significant hurdles, leading to privacy concerns among parents and complicating the researchers' ability to analyze raw usage metadata. We provide actionable interaction design guidelines for developing EMA platforms that prioritize youth agency, ethical practice, and research goals.
\end{abstract}



\maketitle
\input{01_Intro}
\input{02_Background_short}

\input{03_CaseStudy}

\input{04_Method}
\input{05_Findings}
\input{06_Discussion}


\begin{acks}
We thank all participants for their time and contributions. Dr. Wisniewski’s research is supported by the U.S. National Science Foundation under grants \#TI-2550746, \#CNS-2550834, and \#IIS-2550812, and by the William T. Grant Foundation under grant \#187941. Any opinions, findings, and conclusions or recommendations expressed in this material are those of the authors and do not necessarily reflect the views of the research sponsors.
\end{acks}

\balance
\bibliographystyle{ACM-Reference-Format}
\bibliography{References}

\end{document}

%% file: 01_Intro.tex
\section{Introduction}

The IDC community acknowledges how the unique developmental needs of youth shapes the way adolescents experience and negotiate technologies in their everyday lives \cite{minal15support, katie25boundary}. Foundational work by \citet{poole2013interaction} has emphasized that studying adolescents requires methods and research protocols attuned to youth agency, evolving capacities, and power asymmetries between young people and adults. Traditional approaches such as retrospective surveys \cite{barry2017adolescent, chins2024ozzie}, structured interviews \cite{foss2013recruiting}, and adult proxy reports \cite{caddle2023duty} often struggle to capture experiences as they unfold and risk positioning youth as passive subjects rather than active contributors to knowledge production. Conversely, youth-centered participatory methods, which have been championed within the IDC community, center adolescents’ voices, agency, and perspectives through approaches such as youth advisory boards, participatory design, and co-design \cite{samreen2025engage, nelson2025helped, badillo2019stranger}. However, participation alone does not guarantee meaningful or sustained engagement in research \cite{poole2013interaction}, highlighting the need for research approaches that actively nurture and sustain the adolescent’s role as an expert in their own life. 

Building on \citet{poole2013interaction} argument, protocols used to engage adolescents in research must look beyond the mere methodology to the sociotechnical systems shaping participation. To this end, researchers have advocated for inclusive approaches that actively empower adolescent participants rather than just involving them \cite{Oguine2026genai, buckmayer24particpatory}. Ecological momentary assessment (EMA) has emerged as a promising methodological approach that aligns with IDC principles \cite{iverson2012scandanavian-pd} by capturing real-time, in-situ experiences of adolescents while reducing recall burden \cite{shiyko2017feasibility, comulada2015compliance}. EMA has been increasingly adopted in adolescent research to study emotions, behaviors, and contexts as they unfold in daily life. At the same time, EMA introduces new demands, requiring repeated engagement, coordination, and sustained motivation over extended periods \cite{mcintyre2016longitudinal}. The IDC research community is uniquely positioned to address these challenges, given its longstanding focus on designing technologies, methods, and research protocols that center children’s and adolescents’ lived experiences, agency, and power. As such, the design of EMA research platforms is not merely a technical concern, but a core interaction design problem that directly shapes participation, retention, and power dynamics in practice. In this paper, we present a case study based on interviews with a research team to examine how an EMA diary platform, comprising a mobile app for adolescents and a web-based portal for researchers that we designed and developed, shaped adolescent participation, researcher workflows, and relationships between youth and researchers. We identify both the benefits and limitations of this platform in practice and outline design considerations for teen-centered research. Specifically, we ask:

\begin{itemize}
    \item \textbf{RQ1:} \textit{What benefits did the EMA platform provide for participants and researchers?}
    \item \textbf{RQ2:} \textit{What challenges emerged when using the EMA platform in adolescent research?}
    \item \textbf{RQ3:} \textit{What design opportunities could improve the research experience?}
\end{itemize}

To address these questions, we conducted semi-structured interviews with six researchers who used an open-source EMA platform that we designed and developed for our own research. This work makes three contributions to the IDC and broader HCI community. First, we present an open-source EMA research platform designed to support adolescent participation and researcher workflows, and demonstrate its use in a large-scale longitudinal study at an external institution. Second, we provide an empirical account of how EMA research platforms shape participation, engagement, and power dynamics in real-world adolescent research. Third, we outline design implications for building flexible and interdisciplinary research infrastructures that better support adolescents, researchers, and caregivers over time.

%% file: 02_Background_short.tex
\section{Background: EMA Diary Studies with Adolescents}
Ecological momentary assessments (EMA) and diary methods have been widely used to study adolescents’ experiences, from everyday stress \cite{comulada2015compliance} to more sensitive topics like self-harm and suicidal ideation \cite{grist2018acceptability}, because they offer several methodological advantages for research with adolescents. 
By collecting brief responses at multiple moments, often via mobile devices, EMA reduces reliance on retrospective recall and captures experiences closer to when they occur. This approach is particularly well suited to adolescence, as teens’ emotions, social contexts, and daily routines often shift rapidly over short periods of time \cite{shiyko2017feasibility, comulada2015compliance}. Prior work demonstrates that mobile experience sampling can capture fine-grained emotion dynamics in real time, offering more precise insight into adolescents’ everyday experiences than retrospective methods \cite{kang2022understanding, limberger2023assessing}. 

At the same time, EMA introduces significant challenges related to feasibility and sustained participation. Adherence and completion have been consistently identified as central concerns in adolescent EMA research, particularly in longitudinal studies \cite{shiyko2017feasibility, comulada2015compliance}. For example, Shiyko et al. highlight adherence as a key constraint in EMA studies with urban minority youth, while Comulada et al. show that even when compliance is high, it remains contingent on prompt timing, study duration, and fit with adolescents’ daily lives \cite{shiyko2017feasibility, comulada2015compliance}. When protocols do not align with daily routines, engagement can decline, limiting both data quality and participant experience.
To address these challenges, many EMA studies incorporate features intended to support engagement and retention, such as immediate value \cite{webb2021app}, well-being resources \cite{mens2022promoting}, or personalized feedback \cite{dietvorst2024real}.

Given these demands, EMA platforms require careful design to balance research goals, participant experience, and ethical considerations. Within IDC and the broader HCI community, participatory and youth-centered design approaches offer useful strategies for informing the design of research tools by drawing on adolescents’ perspectives and everyday practices. Drawing on participatory work with both teens and researchers, we co-designed and developed a custom EMA diary platform consisting of a mobile app for adolescents and a web-based portal for researchers. Based on the teens' big ideas from their storyboarding exercise, the platform was designed to support longitudinal EMA research while attending to adolescent engagement, ethical considerations, and researcher workflows, extending beyond survey delivery to include features such as progress tracking, incentives, help resources, and oversight mechanisms. This case study examines how research tools shape participation and research work over time.

%% file: 03_CaseStudy.tex
\section{System Overview: The EMA Diary Study Platform}
This section discusses the EMA diary study platform, a daily, real-time system co-designed with adolescents, \edit{where youth input shaped not only interaction features (e.g., diary entries, reminders, and incentives) but also considerations around privacy, trust, and participation in sensitive research contexts (see our prior work for a detailed description of the co-design process and app features \cite{badillo-risky, badillo_30days}).} The tool consists of two components: a mobile app for adolescent participants to submit daily entries, and a web-based portal for researchers to manage participants and review data. \autoref{fig:mobile-app} and \autoref{fig:webtool-tool} show representative screenshots of the two interfaces \cite{badillo_30days}.

\subsection{EMA Mobile Diary App for Teens}
\begin{figure*}[t]
\centering
\includegraphics[width=0.97\textwidth]{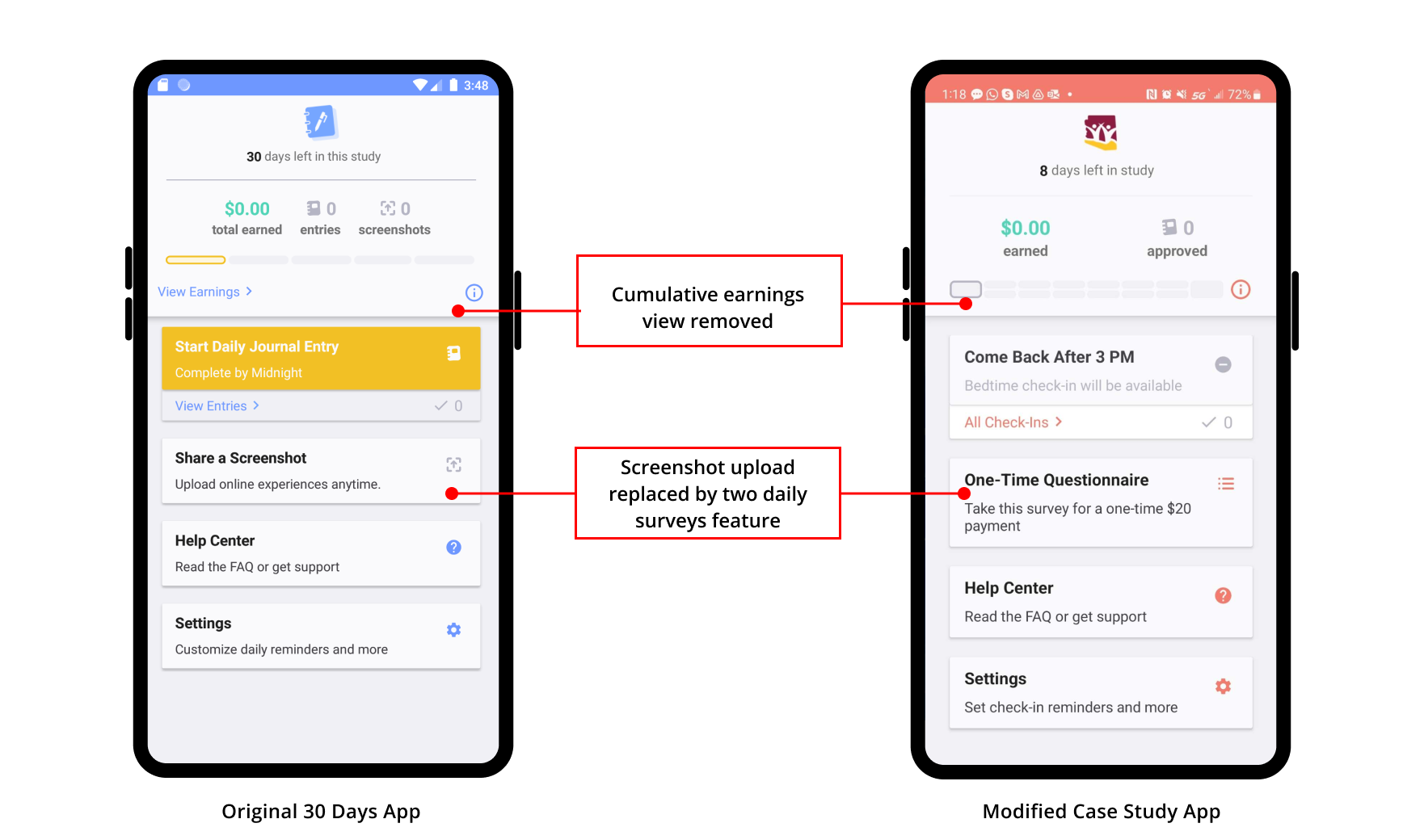}
\caption{Original EMA app (a) and modified case study app (b) key feature differences.}
\label{fig:mobile-app}
\end{figure*}

    The mobile application is a cross-platform app (Android and iOS) designed to support daily diary participation over a 30-day period (see \autoref{fig:mobile-app}a). The key features of the mobile app include:
    \begin{itemize}
        \item \textbf{Dashboard:} Displays study progress, total earnings, and provides access to journal entries, screenshots, submission history, help resources, and app settings.
        \item \textbf{Daily Journal:} Prompts participants daily to complete a survey about their online experiences, after which data are uploaded and the user returns to the dashboard.
        \item \textbf{Earnings:} Shows a breakdown of compensation earned through approved journal entries and screenshots across study periods.
        \item \textbf{Screenshot Upload:} Allows participants to upload screenshots of online interactions at any time, with a required text description accompanying each submission.
        \item \textbf{History:} Enables participants to view submitted journal entries and screenshots, including timestamps, review status, and optional researcher feedback.
        \item \textbf{Help Center:} Provides study information, data collection explanations, support resources, and direct contact options for the research team and external hotlines.
        \item \textbf{Settings:} Allows participants to customize reminder timing, manage a 4-digit app passcode, and sign out of the app.
    \end{itemize}

\subsection{Web Portal for Researchers} 
\begin{figure*}[t]
\centering
\includegraphics[width=0.97\textwidth]{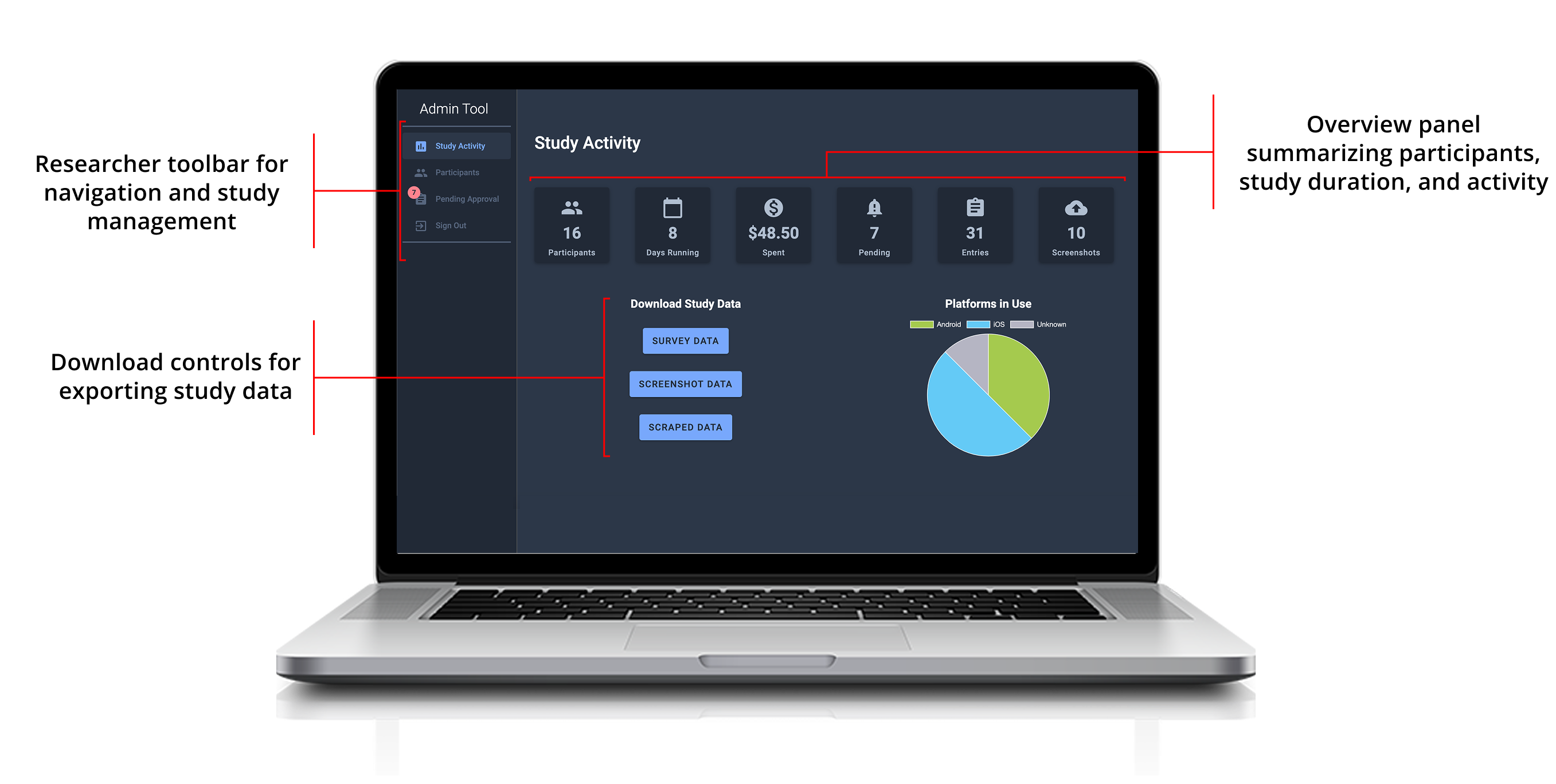}
\caption{Web portal dashboard of the 30-Day EMA diary tool.}
\label{fig:webtool-tool}
\end{figure*}

    The web portal supports researchers oversight of the study (See \autoref{fig:webtool-tool}). And the key features of the web portal include:
    \begin{itemize}
        \item \textbf{Study Activity:} Provides an overview of study status, including participant counts, submission totals, pending reviews, cumulative earnings, device distribution, and data export options.
        \item \textbf{Participants:} Allows researchers to enroll, view, manage, or withdraw participants and access individual profiles with demographics, progress, earnings, and submission records.
        \item \textbf{Submissions:} Displays all journal and screenshot submissions, enabling researchers to review, approve or reject entries, provide feedback to participants, and add internal notes.
    \end{itemize}

\subsection{Case Study: EMA Diary Twin Study}
The EMA diary platform was deployed by the Arizona Twin Project research team conducting longitudinal psychological studies with adolescent participants \cite{ASU2026twin-project}. They integrated the tool into their research workflow to enable direct, daily self-reports from teens, rather than relying on parent-reported diaries. Depending on the study wave, participants completed app-based check-ins over an extended period. These entries captured information about daily activities, sleep, emotions, stress, social support, and electronic media use. To align the tool with their study protocol, the research team replaced some features of the original mobile app (\autoref{fig:mobile-app}a). Most notably, the screenshot upload functionality (2) was replaced with a questionnaire-based feature that prompted participants to complete two daily surveys. As shown in \autoref{fig:mobile-app}b,  a dual progress bar was introduced to indicate completion of both daily entries. In addition, the option for participants to view cumulative earnings was removed to better align with the study’s incentive structure.



%% file: 04_Method.tex
\section{Interview Study Methods}

We conducted semi-structured interviews with members of the Arizona Twin project that integrated a 30-day EMA diary platform into their longitudinal study workflow. Interviews focused on researchers’ experiences using the mobile app and web-based portal, including how these tools supported (or hindered) participation, coordination, and study management. The interview guide covered three areas: researchers’ roles and prior experience, use of the EMA app and web portal in practice, and reflections on challenges and potential design improvements. Follow-up questions were used to probe emerging themes and clarify responses.

\subsection{Data Collection and Participant Recruitment}
We obtained IRB approval from our institutional review board at the University of Notre Dame before beginning data collection. Interviews were conducted over Zoom to accommodate participants’ schedules. Prior to each interview, participants received an information sheet and provided verbal consent to participate and to be video-recorded. Recruitment occurred over one month (February to March 2025). We emailed 13 team members and coordinated with the project's Primary Investigator (PI) to ensure outreach was distributed across the group. In total, we conducted six interviews, each lasting an average of 40 minutes, resulting in approximately 4 hours and 22 minutes of recorded material. Five participants identified as female and one as male. The group included one PI, two graduate research assistants, two undergraduate research assistants, and one administrative staff member who had previously worked as a research assistant. All participants had at least one year of experience on the research team.

Before each interview, participants provided verbal consent and agreed to be video recorded. Recruitment took place between February and March 2025 through email outreach to team members with permission from the PI. We conducted six interviews, each averaging 40 minutes, totaling approximately 4 hours and 22 minutes of recordings. Participants included one PI, two graduate research assistants, two undergraduate research assistants, and one administrative staff member; five identified as female and one as male. All participants had at least one year of experience on the research team. We obtained IRB approval prior to data collection. 

\begin{table*}[t]
\centering
\small
\caption{Participant Profiles}
\label{tab:participant_profiles}
\begin{tabular}{l l l c}
\toprule
\textbf{Pseudonym} & \textbf{Academic Level} & \textbf{Team Role} & \textbf{Years on Project} \\
\midrule
Cari & Graduate RA & On-call Team Lead & 3 years \\
Valery & Graduate RA & Risk-Assessment Team Lead & 5 years \\
Wendy & Undergraduate RA & On-call Team Member & 2 years \\
Daniel & Undergraduate RA & On-call Team Member & 1 year \\
Sophie & Administrative Staff & Staff Member & 2 years \\
Nichole & Project Investigator (PI) & Project Coordinator & 23 years \\
\bottomrule
\end{tabular}
\end{table*}

\subsection{Qualitative Analysis Approach}
To analyze the interview data, we conducted a reflexive thematic analysis \cite{braun2019reflecting} of the transcripts. This approach allowed us to iteratively identify and interpret patterns in how researchers in our case study experienced and used the EMA diary platform. The first and second authors led the coding, with input from the remaining authors. We began by reading each transcript multiple times to build familiarity with the data. We then conducted open coding, generating and refining preliminary codes into conceptual groupings. These groupings were iteratively developed into subthemes and further grouped into themes through consultations with the coauthors during regular weekly meetings. Since we worked together closely during the analysis phase and followed a consensus-coding approach, calculating inter-rater reliability was not necessary \cite{mcdonald2019reliability}.

%% file: 05_Findings.tex
\section{Findings: Lessons Learned}
The following subsections describe how the EMA platform facilitated adolescent engagement and streamlined research administration, while also surfacing sociotechnical drawbacks and design opportunities that shaped participation, data quality, and research workflows.

\subsection{EMA Platform Facilitated Adolescent Engagement and Streamlined Research Administration (RQ1)}
The following subsections detail how the EMA platform supported adolescent engagement through seamless onboarding, centralized task management, and reminders, while simultaneously enabling researchers to manage risk and streamline study administration.

\subsubsection{EMA app onboarding process was seamless, not requiring a steep learning curve.}
Researchers consistently described onboarding as seamless due to the app’s intuitive and simple design, which reduced the initial learning burden for adolescents. This ease of onboarding was attributed to interaction patterns that aligned with adolescents’ existing experiences using mobile applications, allowing participants to quickly understand how to navigate the app and begin completing study tasks without extensive explanation. Rather than relying on prior technical expertise or repeated instruction, the interface itself communicated how to proceed, making onboarding a low-effort entry point into the study.
\begin{quote}
    \textit{`` We will have a training session to demonstrate how to use it. However, they already have a basic understanding of how to use it even before the training starts, simply by logging in; it is really that intuitive. For these reasons, I would say it is very beneficial. Having an app that is user-friendly for the [participants].''} - Daniel
\end{quote}

\subsubsection{EMA app dashboard facilitated engagement by centralizing daily tasks}
Researchers further emphasized that the app’s centralized dashboard played an important role in this process by consolidating study activities, timing information, and progress indicators in a single location. This reduced the need for participants to navigate multiple screens or seek clarification about next steps, and in turn lowered the amount of researcher intervention required during early participation. When issues did arise, the simplicity of the app also supported faster troubleshooting, enabling both researchers and participants to identify and resolve minor problems before they escalated into missed check-ins or disengagement.

Beyond lowering entry barriers, researchers emphasized that the app supported sustained participation by reshaping how adolescents experienced engagement in the study. Rather than framing participation as an obligation requiring repeated reminders from staff, the app made progress and rewards visible, immediate, and motivating. Researchers highlighted the gamified incentive features as a primary driver of engagement, noting that seeing earnings accumulate helped connect daily participation with tangible outcomes. As one researcher observed:
\begin{quote}
    \textit{``One major feature that has been really helpful is showing how much the [participants] are earning as they complete tasks[...] it's kind of fun for the [participants] to see their earnings, which is good for motivation and retention as well.''} -  Sophie
\end{quote}
Researchers also described how teens used the EMA app in ways that went beyond simply completing required tasks. In some cases, adolescent participants treated the app as a private space for reflection, given the open-ended survey tasks. Rather than feeling like a formal research instrument, the mobile interface supported a more informal, self-directed mode of expression. Researchers viewed this as beneficial for data collection, noting that the diary-like interaction encouraged fuller and more candid responses, especially in open-ended entries. Some of the researchers observed that adolescents seemed more comfortable sharing their experiences. For instance:
\begin{quote}
    \textit{``I also think having the app is from what I've noticed. Sometimes, in some cases, the [participants] have used it as a way to run open-ended responses. They actually feel a lot more comfortable sharing more information because it just kind of feels like, almost like a Mini diary that they're typing into like on their phone. And so I think for data collection purposes, it's really nice for more like objective data collection''} - Sophie
\end{quote}
This shows that the EMA platform supported engagement by offering a familiar, diary-like interface that made it easier for participants to share over time.

\subsubsection{EMA app reminders improved sustained engagement overtime.}
Engagement was further supported by reminders that ensured timely notification of participants to complete daily study tasks. By notifying participants when check-ins opened and closed, the app reduced uncertainty around timing and made it easier to complete tasks in the moment rather than postponing and forgetting. Importantly, researchers described how embedding incentives and reminders within the app shifted responsibility for participation away from staff and toward adolescents themselves. This reduced the need for repeated prompting through emails or messages, which researchers viewed as potentially intrusive over time, and instead supported adolescents’ ability to self-regulate their participation, altering the power dynamics of engagement.

\begin{quote}
    \textit{``We just sent them an absurd number of text messages, to which most of the time they don't respond. Yeah, I think it's a lot more effort on our end. [...], they just get a ding on their phone reminding them. ''} - Valery
\end{quote}
This indicates that reminders embedded in the EMA app reduced forgetfulness and researcher effort, thereby supporting sustained adolescent participation over time.

\subsubsection{Researcher web portal supported risk mitigation and study administration.}
Researchers described the web portal as essential for oversight and coordination in a complex study. A key benefit was timely risk monitoring, particularly through prominent access to open-ended responses that often contained safety-relevant information. Being “on call” to assess indicators such as suicidality, researchers emphasized that the portal made this information easy to locate and act upon.

\begin{quote}
     \textit{``One of the main things we do while on call is monitor risk tracking. We focus on indicators of suicidality, homicidality, and similar factors; I find that information tends to be relatively easy to locate.''} - Valery
 \end{quote}

Beyond risk assessment, the portal also reduced cognitive and administrative workload by providing a clear study overview. Researchers could quickly assess completion status, identify missing check-ins or uploads, and prioritize follow-up without reviewing every individual response. Compared to survey platforms alone, this centralized view substantially reduced manual tracking and coordination, especially for research assistants.

\subsection{Power Dynamics and Sociotechnical Constraints Constrained EMA App Use in Adolescent Research (RQ2)}
The following subthemes illustrate how power differentials among adolescents versus parents and researchers shaped EMA app use, surfacing tensions around authority, access, and control that influenced privacy decisions, participation, and the interpretation of study data.

\subsubsection{Parental influence on data collection and privacy decisions.}
One recurring source of tension involved parental concerns about data collection and privacy, particularly given the app’s collection of background usage data. Several researchers reported instances where parents questioned why certain permissions, such as location services, were required and what types of data were being collected, often reflecting broader concerns about surveillance and oversight of their children’s online activity. In response, researchers frequently needed to provide additional explanations about the purpose of these data and the safeguards in place to protect participant information. Although these concerns rarely prevented participation, they introduced moments of hesitation that required reassurance and trust-building. As one researcher explained:
\begin{quote}
     \textit{``When we set up the app, there are questions about it. Oh, why do we have location services on? Why are you asking for certain data? We will also explain why we are doing these things, while ensuring that all their information is secure and protected.''} - Daniel
 \end{quote}
A related tension emerged around the app’s visible incentive structure. While researchers viewed progress tracking and compensation visibility as effective motivators for adolescents, some parents expressed discomfort with their teens seeing how much money they were earning. In these cases, parents’ expectations about financial control and disclosure conflicted with the app’s default design, which assumed that transparency around incentives would be universally acceptable. Researchers described instances where families requested that compensation be directed to parents rather than to the adolescents themselves, or that information about earnings not be shown to the teen participants at all. These situations required researchers to navigate differing family preferences and, in some cases, rely on external explanations or workarounds to align the app’s behavior with parental expectations.
\begin{quote}
    \textit{ ``Whether the money that they make goes to the family itself or goes to the [participants] ... occasionally, the parents will say, we don't want any money to go to the [participants]. You know the money will go to us.''} - Cari
 \end{quote}

\subsubsection{Sociotechnical constraints limited adolescents’ access and ability to participate}
In addition to these expectation-based tensions, researchers and teen participants encountered frustrations stemming from system constraints related to access, connectivity, and platform design. One recurring challenge involved onboarding dependencies on email access and temporary password delivery. Researchers reported cases where participants were unable to retrieve passwords due to full email inboxes, forgotten credentials, or limited internet access, particularly in rural areas with unreliable connectivity. Although these issues were not directly caused by the app itself, they disrupted onboarding and delayed participation.
 \begin{quote}
     \textit{``Some participants live in very rural areas where they don't have easily accessible Wi-Fi or any other kind of internet connection. That makes it really challenging for them to use the app.''} - Cari
 \end{quote}

Meanwhile, researchers emphasized how limited system flexibility concentrated troubleshooting responsibility on the research team. Once participant accounts were created and temporary passwords issued, correcting even routine administrative errors often required researchers to withdraw and re-register participants, increasing their workload. \edit{This introduced additional dependencies on researcher support, which they perceived as shaping participation dynamics and limiting adolescents’ autonomy in practice.}

\begin{quote}
    \textit{``We just sent them an absurd number of text messages, to which most of the time they don't respond. Yeah, I think it's a lot more effort on our end. [...], they just get a ding on their phone reminding them. ''} - Valery
\end{quote}

Researchers also described technical issues and resource use, including software bugs, accidental app closures, and increased battery consumption associated with background screen-time collection that affected participant compliance and study continuity. To mitigate these challenges, researchers often preemptively informed participants about potential battery drain to manage expectations and sustain engagement.


\subsection{Design Opportunities to Better Support Research Workflows (RQ3)}
Across interviews, researchers reflected on design opportunities that could better align the system with the realities of longitudinal research workflows. These opportunities were often articulated in response to challenges encountered during data collection, monitoring, and analysis, and pointed toward ways the platform could evolve beyond its current capabilities. 
\subsubsection{Support integration with wearable devices.}
One frequently discussed opportunity involved integrating data from wearable devices to complement self-reported and usage data with physiological measures, such as heart rate or skin conductance. Researchers viewed such integration as a way to enrich contextual understanding of adolescents’ daily experiences while reducing reliance on retrospective reporting. One of the researchers suggested,
\begin{quote}
    \textit{``So how can we integrate into the app communication with perhaps something like wearable devices? Right? So we're gathering real-time information on, you know, heart rate or on skin conductance''} -Nichole.
\end{quote}
 This extension was framed not as a replacement for EMA, but as a means of triangulating data sources to better capture moments of stress, arousal, or disengagement as they occur. 

\subsubsection{Better data structure and visualization to support analysis and sensemaking.}
Researchers identified opportunities to improve data structure and visualization to reduce analytic workload and support faster sensemaking. For instance, they complained that exported data were provided as individual files per participant rather than as a consolidated dataset, requiring extensive preprocessing and increasing opportunities for error. 

\begin{quote}
     \textit{``The other issue is that we receive one Excel sheet per participant. It would be great for us to have one large data file for all participants to analyze.''} -Valery
\end{quote}

In addition, the lack of aggregated views, flexible dashboards, and configurable summaries within the app and administrative portal limited researchers’ ability to identify patterns in real time, particularly during on-call monitoring. Researchers emphasized the value of more pre-processed and visualized data to support timely interpretation rather than relying on post hoc analysis.

\subsubsection{Mobile support for research team.}
Finally, researchers highlighted the absence of a mobile-accessible interface for research staff as a limitation. Research activities such as risk monitoring and participant support often occurred alongside other daily responsibilities, requiring staff to check submissions, assess alerts, and coordinate responses while away from their desks. Given the time-sensitive nature of this work, a mobile version of the administrative tool was viewed as essential for reviewing submissions, tracking study progress, and responding to issues in situ rather than deferring action until later.

\begin{quote}
    \textit{``But if there were an Admin tool app for us to use, we could easily navigate things on our phones. I think a lot of our colleagues actually do their work from their phones rather than their computers ''} -Daniel.
\end{quote}
These suggestions highlight the need for research tools that support both participant engagement and the realities of collaborative, time-sensitive research work.

%% file: 06_Discussion.tex
\section{Discussion}


Table \ref{tab:ema_summary} summarizes the benefits and drawbacks identified across the EMA mobile app and researcher web portal. The following sections discuss how the EMA platform’s design shaped adolescent participation, engagement, and power dynamics, and consider the implications of these findings for designing research tools in adolescent studies.

\begin{table*}[t]
\centering
\small
\caption{Summary of Benefits and Drawbacks Identified for the EMA Platform}
\label{tab:ema_summary}
\begin{tabular}{p{2.6cm} p{5.0cm} p{5.0cm}}
\toprule
\textbf{Platform} & \textbf{Benefits} & \textbf{Drawbacks / Challenges} \\
\midrule
\raggedright\textbf{EMA Mobile App} &
Intuitive mobile interface supported quick onboarding; reminders and progress indicators sustained engagement; visible incentives reduced researcher prompting. &
Parental concerns about data collection and incentives; uneven device and connectivity access; technical issues including battery drain and platform-specific data limits. \\
\midrule
\raggedright\textbf{Researcher Web Portal} &
Centralized oversight supported participation tracking and risk monitoring; reduced manual coordination. &
Rigid account management increased administrative effort; fragmented exports required preprocessing; limited visualization hindered real-time sensemaking. \\
\midrule
\raggedright\textbf{Platform Overall} &
Supported longitudinal, in-situ data collection; shifted participation responsibility toward adolescents. &
Ongoing researcher mediation required; lack of mobile access limited responsiveness during on-call work. \\
\bottomrule
\end{tabular}
\end{table*}

\subsection{Designing to Facilitate Participation in Adolescent Research}
Our findings show that EMA platforms can support adolescent onboarding, engagement, and sustained participation through concrete design features rather than outreach or motivation alone. Prior work has identified recruitment and retention as persistent challenges in adolescent research, often attributing difficulties to adolescents’ perceived disinterest \cite{Preston2019May}, limited capacity \cite{Bradbury-Jones2018Oct, Warraitch2024Aug}, parental gatekeeping \cite{Jorgensen2019May}, or consent and access constraints \cite{Cullen2020Sep, jong2023recruitment, wilson2020rapid}. Our results extend this literature by showing that participation is also shaped by how adolescents encounter research tools at entry. In this study, onboarding was supported by an interface that aligned with adolescents’ everyday mobile practices, allowing participants to navigate the app with minimal instruction \cite{shiyko2017feasibility, poole2013interaction}. This aligns with prior findings that adolescents are more likely to engage with systems that feel familiar and legible rather than overtly institutional \cite{murray2024engaging, foss2013recruiting}. However, access to these benefits was uneven, as parental concerns about data collection and disparities in device access and connectivity constrained participation for some families \cite{oguine2025inclusion, menberu2024technology, sharma2024whose}. 

Meanwhile, sustained participation was supported through in-app reminders, progress indicators, and incentives embedded directly in the app. Timely notifications reduced missed check-ins, while visible progress and compensation helped adolescents connect daily participation to immediate outcomes. Prior work has shown that incentives \cite{poole2013interaction}, parental encouragement \cite{Robbins2011Sep}, curiosity \cite{brawner2013attitudes}, and social context \cite{Crane2017Feb} influence retention. Our findings add that how these elements are implemented matters. Progress tracking and non-monetary feedback supported engagement even when parents limited adolescents’ visibility into compensation, suggesting that participation can be sustained without relying solely on financial incentives \cite{Warraitch2024adress-barriers, Preston2019May}. Based on these findings, EMA tools for adolescent research should prioritize familiar interaction patterns during onboarding, embed engagement support within the system rather than relying on researcher prompting, and provide flexible mechanisms for motivation that can adapt to parental constraints and access differences.

\subsection{Negotiating Power and Agency in Adolescent Research}
Our work contributes to ongoing discussions about power imbalances in adolescent research \cite{poole2013interaction, Warraitch2024adress-barriers}. Prior work has shown that adolescents occupy a structurally unequal position relative to adult researchers, parents, and institutions, even when participation is voluntary and assent procedures are followed \cite{Sellars2021Jul, Cullen2020Sep}. While assent is intended to support ethical understanding and willingness, it does not fully account for the influence of adult authority, parental expectations, or perceived obligation \cite{Robbins2011Sep}. In this study, the design of the EMA tools shaped how power was exercised in practice. Researchers used the app to reduce direct interpersonal pressure on adolescents by embedding reminders, incentives, and progress tracking into the system. This allowed researchers to step back from repeated prompting and gave adolescents more control over when and how they participated. This aligns with ongoing and prior work calling for systems that support youth autonomy by reducing adult enforcement \cite{wang202312, Warraitch2024adress-barriers, Oguine2026genai, Preston2019May}, and shows how these principles can be implemented through design rather than relying only on researcher behavior. 

At the same time, adolescents’ autonomy was constrained by parental involvement and technical limitations that required researcher intervention. Parental concerns about data collection and incentive visibility often conflicted with design assumptions centered on transparency and youth agency \cite{poole2013interaction, Jorgensen2019May}. These concerns were frequently shaped by family norms and material conditions, positioning parents as decision-makers around participation and compensation. Researchers also served as intermediaries when system rigidity made it difficult for adolescents to resolve onboarding or account issues independently, reinforcing the research team’s authority and responsibility for participation. Like prior work \cite{Warraitch2024adress-barriers, cluver2021power}, our findings show that power in adolescent research is distributed across adolescents, parents, researchers, and the systems that mediate participation. While the EMA platform supported adolescent engagement, they also reproduced dependencies that limited adolescents’ autonomy. Sociotechnical tools designed to support teen-engaged research should better account for these dynamics by creating pathways for adolescents to exercise greater agency, while reducing reliance on parental oversight and researcher troubleshooting.

\subsection{Implications for IDC Research and Beyond}
Our findings suggest several implications for how IDC researchers design research platforms for adolescent research. These guidelines focus on supporting adolescent agency, managing power dynamics, and aligning research infrastructures with interdisciplinary workflows.

\begin{itemize}

    \item \textbf{Design for adolescent agency within shared power structures.} Platforms should support adolescents’ ability to manage participation through clear feedback, self-regulation features, and meaningful control, while remaining responsive to parental authority and researcher oversight. Adjustable levels of visibility and control can help shift power toward adolescents without ignoring institutional constraints.

    \item \textbf{Design tools for interdisciplinary research teams, not only participants.} Research platforms should support collaboration across disciplines by enabling coordination, shared sensemaking, and risk monitoring, particularly in studies that connect adolescents’ online behaviors with offline outcomes such as mental health.

    \item \textbf{Treat researcher mediation as a design signal.} Points where researchers must intervene, such as clarifying data practices or correcting administrative errors, should be addressed through flexible, transparent system design to reduce hidden labor and support real-world research workflows.

    \item \textbf{Support multiple analytic workflows.} Research platforms should provide both programmable data access and clean, consolidated datasets to accommodate different disciplinary analysis practices and reduce translation work.

\end{itemize}

\noindent These guidelines emphasize that designing EMA platforms for adolescent research requires attending not only to engagement and data collection, but also to how participation, power, and research labor are structured in practice.

\section{Limitation and Future Work}
This study focuses on researchers’ experiences using an EMA platform with adolescents, which allowed us to examine how tool design shaped participation and research coordination. \edit{Notably, participants in this study drew on their broader experience conducting multiple longitudinal studies with adolescents, allowing insights to extend beyond a single deployment.} However, it does not capture how adolescents or parents experienced the app or how they understood participation, control, and trust. Future work should include adolescents’ and parents’ perspectives to better understand how participation and power is negotiated in practice. In addition, this study examines a single EMA platform used in one research context, which may limit how broadly these findings apply. Future studies should examine EMA platforms across different populations of teens, study designs, and infrastructural settings. Building on this work, future research should explore how EMA systems can better support adolescent agency, accommodate parental involvement, and reduce the coordination and troubleshooting work required of research teams.

\section{Conclusion}
This paper highlights how the design of EMA research platforms can influence who is able to participate, how participation is sustained, and how responsibility and control are distributed among adolescents, parents, and researchers. As EMA methods continue to expand in youth research, there is a growing need to design tools that do not simply enable data collection, but actively support adolescents’ autonomy while remaining responsive to parental involvement and interdisciplinary research work. Given that design is an intervention, IDC work can help ensure that research technologies better support adolescent participation in research while accounting for power dynamics and the practical realities of conducting teen-engaged research.

\section{Selection and Participation of Children}
This paper does not include adolescent participants. While the EMA platform was deployed in a parent study involving adolescents, the research reported here is based exclusively on interviews with adult members of the research team who designed, deployed, and managed the platform, and was approved by our institutional review board (IRB). Adolescents are discussed only in relation to the case study, where they participated under parental consent and adolescent assent by completing app-based EMA check-ins. All findings therefore reflect researchers’ perspectives on how the platform shaped youth participation and engagement.